\documentclass[12pt]{iopart}
\usepackage[LGR,T1]{fontenc}
\usepackage{graphics, graphicx,color,soul}
\usepackage[dvips]{epsfig}
\DeclareGraphicsExtensions{.jpg,.pdf,.mps,.png}
\definecolor{bleuclair}{rgb}{0.7, 0.7, 1.0}
\sethlcolor{yellow}

\begin{document}

\title{A modern Fizeau experiment for education and outreach purposes}

\author{O. Morizot, A. Sellé, S. Ferri, D. Guyomarc'h, JM. Laugier, M. Knoop}

\address{Physique
des Interactions Ioniques et Mol\'eculaires, UMR 6633 - CNRS et Aix-Marseille Universit\'e,
 Centre de Saint J\'er\^ome, Case C21,
13397 Marseille Cedex 20, France}
\ead{Martina.Knoop@univ-provence.fr}
\begin{abstract}
On the occasion of the laser's 50th
anniversary,  we performed a modern Fizeau experiment,
measuring the speed of light with a laser beam passing over the city
centre of Marseille. For a round trip distance of almost five kilometers, the
measurement has reached an uncertainty of about 10$^{-4}$, mainly
due to atmospheric fluctuations. We present the experimental and
pedagogical challenges of this brilliant outreach experiment.
\end{abstract}

\submitto{\EJP}
\maketitle

All throughout 2010, many experiments and activities were proposed  to celebrate
the 50th anniversary of the laser (LASERFEST). Locally, we decided to propose a visible and spectacular
event that would appeal to a wide public and shine light on the laser. The idea was to send a brilliant laser beam in the night sky over the city centre of Marseille. Furthermore we wanted to use this fascinating beam to realise a remarkable experiment, that could attract students, scientists and curious minds in general, and allow us to communicate about science, research and the position of a researcher in an unusual way.
The measurement of the speed of light stroke us as an amazing example of how limits excite the curiosity of the
layperson and the specialist.

In this manuscript we describe our experiment to measure the speed
of light with a modified Fizeau set-up. This choice has been
inspired by the spectacular and successful realization of colleagues in Paris during the World Year
of Physics in 2005~\cite{amp2005}. In this manuscript, we will first briefly review Fizeau's
historic measurement in section~\ref{sec.fiz}, before describing our own
experimental set-up in section~\ref{sec.exp}. At some point, we tried to reproduce the historical set-up
by using a rotating cogwheel, and this will be discussed in
section~\ref{sec.roue}. The results, systematic
effects and precision of our measurement will then be discussed
in section~\ref{sec.results}. Finally, a summary of the
accompanying outreach activities that we have proposed to the wide
public will be given in section~\ref{sec.outreach}.

\section{Fizeau's measurements}
\label{sec.fiz}
In 1849 Hippolyte Fizeau~\cite{fiz1849} was the first to measure the speed of light via a terrestrial experiment. His method measured the time needed for light to travel to a mirror at a known distance and return. For that purpose he designed a set-up where a collimated beam emitted by a limelight passes through a half-mirror and a rotating cogwheel, is then reflected back by a mirror situated some 8.633 kilometers away, passes (or not) through the cogwheel again, and is reflected by the half-mirror into a monocular.

On the way from the source to the mirror, the beam passes thus through a rotating cogwheel. At a low rotation rate, the light passes through the same blank of the wheel on the way out and on the way back. But with increasing rotation rate, a higher and higher percentage of the transmitted light  is cut on its way back by the incoming tooth of the wheel, resulting in a decreasing light intensity collected in the monocular. Total extinction of the returning light is reached when the time duration of the open gate corresponds exactly to the duration of the round-trip, such that the light that has gone through finds the gate closed when it returns. Knowing the precise distance $d$ between the wheel and the mirror, the number of teeth $N$ of the wheel, and its rotation rate $\omega$ (expressed in radians per second), the speed of light $c_l$ in air can be deduced to be

\begin{equation}
c_l = 2d \times 2N \times f_c
\label{c}
\end{equation}
where $f_c = N\omega/2\pi$ is the frequency at which the beam is effectively stopped. Obviously, if one increases further the rotating speed of the wheel, light will appear again as the returning light will start passing through the gap situated right after the one it has passed on its way out.

Using this method with the cogwheel placed in Montmartre and the reflector in Suresnes, Fizeau obtained a value of $c_l=$~315,000~km/s, limited by the precision of his measurement of $\omega$, but yet better than any measurement realized before.

\section{Experimental set-up}
\label{sec.exp}

\begin{figure}
\begin{center}
\includegraphics[width=130mm]{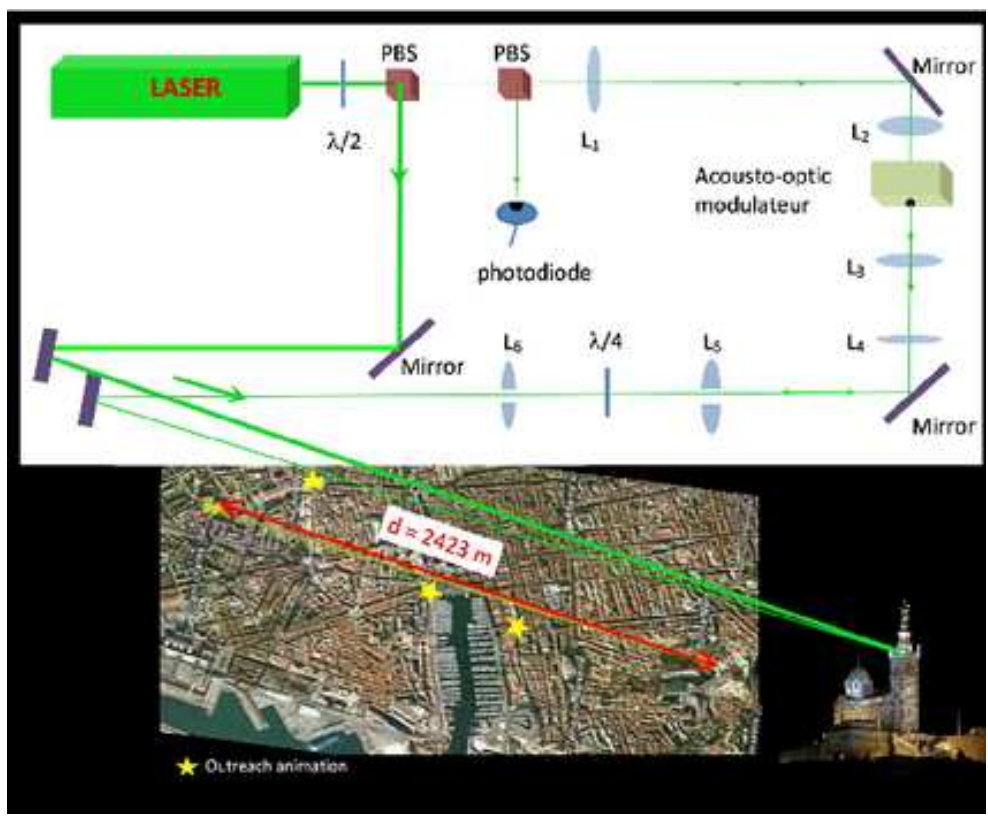}
\caption{Experimental set-up of the modern Fizeau experiment realized in Marseille in June 2010, PBS: polarization beam splitter, L: lens, $\lambda$/4; $\lambda$/2: quarter-; half-wave plate. Detailed description is given in the text, section~\ref{sec.exp}.}\label{fig.set-up}
\end{center}

\end{figure}

The experiment we have set up is deeply inspired by Fizeau's, and its principle is exactly the same. Yet one will notice a few differences to the original experimental set-up along the description below. Our experimental set-up, simple and compact, is sketched on figure~\ref{fig.set-up}. Everything fits on a 60~cm $\times$ 90~cm optical breadboard.

First, our light source was obviously a laser, since the experiment was built in the frame of the 50th anniversary of the laser. We used a  frequency-doubled cw Nd:Yag laser, at a wavelength of 532~nm and an optical output power of 10~W. Due to its high power and its green color, this laser is very visible to the human eye.

Light emitted by the laser first crosses  a half-wave plate and a polarization beamsplitter cube (PBS), that will allow to split the light in two beams of adjustable relative intensities.  A second PBS is aligned with the first one, and therefore transmits all the outcoming light. This PBS serves to totally reflect  the returning light towards the photodiode that is used as the detector for light extinction (instead of a monocular and an eye in Fizeau's case). Note that for the returning light to be reflected by the PBS, its polarization must be orthogonal to the one of the outgoing light. That is the role of the quarter-wave plate (QWP) situated in the lower part of the scheme, that will be crossed twice by the light and thus act as half-wave plate for the returning light. This QWP is placed at the end of the optical bench in order not to affect the parasite reflections on the other optical elements and prevent them from being reflected  to the detector.

The major difference with Fizeau's set-up is certainly the use of an acousto-optic modulator (AOM) instead of a rotating cogwheel in order to chop the light. The main argument for this is certainly the difficulty of realizing a smooth, stable and safe mechanical system rotating at the expected frequency. A detailed discussion of this point is developed in section~\ref{sec.roue}.

The amplitude of the 80~MHz sinusoidal signal driving the the AOM is modulated by a square signal oscillating between 0~V and $V_{max}$ at a frequency $f$, switching on and off the diffraction by the crystal. We defined the main optical path of the experiment to be the first order of diffracted light (+1), as order 0 is never completely extinguished. All other orders of diffraction are eliminated by a set of diaphragms and light traps. We are then left with a single collimated laser beam blinking at frequency $f$, exactly like in Fizeau's case. Note that the operation of an AOM is symmetric and will act in an identical way on the returning light.
Only returning light that encounters the AOM in state "on",  will
be diffracted back and return exactly along its incoming path, to finally reach the detector. In any intermediate case where the returning light pulse is not exactly synchronized with the AOM, only part of the pulse will reach the detector and the rest will be cut off, resulting in the type of signal represented in the upper part of figure~\ref{fig.signal}. This behavior follows equation~\ref{c}, where $f_c$ is the frequency value at which light completely disappears from the detector.

The AOM we used was not adapted to the working wavelength, with a diffraction efficiency of below 40~$\%$ in the first order, and thus an important loss factor. In order to dispose of a very visible beam in the night sky for outreach purposes, we should have had several times higher power at the entrance of the AOM. However, even though on a normal day one would recollect  only 0.02~$\%$ of the outgoing light power on the photodetector, it has turned out that the measurement can be made with as little as 100~mW outgoing laser power. This is why we have split the light in two beams as mentioned at the beginning of this section (see also figure~\ref{fig.set-up}); resulting in one high power beam, completely unaffected by any optical element and mobile for show effects, and one beam dedicated to the measurement were we would not inject more than 2~W, which is sufficiently low power to protect our optical components.

All other lenses on the optical bench help to reshape the returning beam and make it fit through the opening of the AOM without to much losses. One can notice for instance that the final pair of lenses (L5 and L6), set as a telescope, are drilled along their optical axis in order not to act on the outgoing beam and only to affect the diameter of the returning beam, so that it is not clipped by the AOM.

The reflector has been mounted on the most prominent and famous building of Marseille: the church of Notre-Dame de la Garde sitting on a hilltop over the city centre. A 2-inch diameter corner cube was mounted on the upper terrace (no public access !) of its main tower. This solution, contrary to a simple mirror, allows to collect a fair amount of reflected light without the need to align the reflector very accurately at the setting of the experiment, nor to have to correct its orientation for any accidental displacement or misalignment of the light beam. Note that with the normal divergence of the laser beam (< 0.5~mrad), its diameter on the tower is of approximately 2.5~m. Even with a beam perfectly centered on the corner cube, the amount of reflected light will then be less than 11~$\%$. Furthermore, the light reflected by the corner cube keeps on diverging with the same angle on the returning path, which means that back on the experimental set-up, the light beam will be 4-inches wide. It is therefore important that the ultimate mirrors and lenses of the experiment are as wide as possible in order to collect as much returning light as possible.

The exact distance from the AOM situated in a room of the university site Saint-Charles  to the reflector installed on top of Notre Dame de la Garde has been measured as $d$~=~2423.175~m~$\pm$~3.25~cm. This measurement was made by students of the  French National School of Geographic Sciences (ENSG) using in a first step laser telemetry to evaluate the distance between both buildings and then triangulation to  precisely determine the distance with respect to the exact position of the AOM.

\section{The rotating cogwheel}
\label{sec.roue} Fizeau was a fine experimentalist and his rotating
tooth-wheel with its mechanics were an ingenious system \cite{cornu}. We made several
attempts to design a mechanical device for beam-chopping with a precision of the order of a few percent. For the
described 5 km round trip of the laser beam, the chopping frequency must be as high as 30~kHz.

\begin{figure}
\begin{center}
\includegraphics[width=85mm]{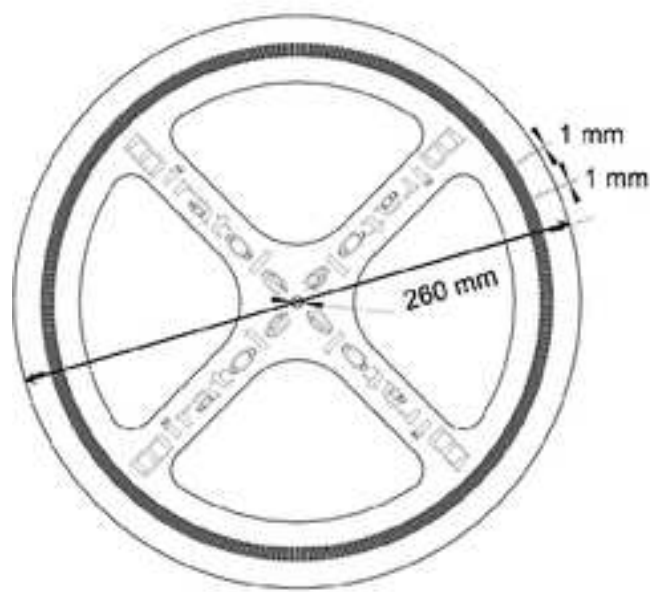}
\caption{Geometry of the rotating cog-wheel, all dimensions in mm.}
    \label{fig_roue}
\end{center}
\end{figure}

One solution to reduce the rotating rate of the wheel is to increase the number of its teeth. The designed wheel has a diameter of 260~mm, is made of 1~mm-thick aluminium (figure~\ref{fig_roue}), and presents 360 holes that were laser-cut (see fabrication movie on~\cite{roue}). These holes are 0.5~mm wide in
order to easily focus the laser beam without clipping, and the symmetry of holes and teeth is better than 5~\%. The wheel then only needs to turn at a frequency of approximately 85~Hz. Yet, this rotation should be
uniform, and must be variable in a range of about $\pm$~15~Hz, in
order to be able to measure a minimum of the reflected light.
  
In order to decrease the weight of the wheel, its inner part was hollowed to a maximum extent. Different categories of motors from model equipment have been tested, lacking either torque to drive the 150-grams wheel, or frequency stability in the rotation. The final solution came from the use of an electric drill engine that yet presented the flaw of being uncomfortably noisy.

Another major difficulty is the enormous amount of light scattered all over the room and onto the detector when the beam is cut by a tooth of the chopper. We mounted the wheel in a black enclosure in order to reduce these reflections, but still we could not manage to reach a reasonable signal-to-noise ratio of the detected signal in the short time frame of the experiment, which finally did not allow us to make a significative measurement using mechanical chopping of the beam by a toothed wheel.

\section{Results and discussion}
\label{sec.results}

\begin{figure}
\begin{center}
\includegraphics[width=100mm]{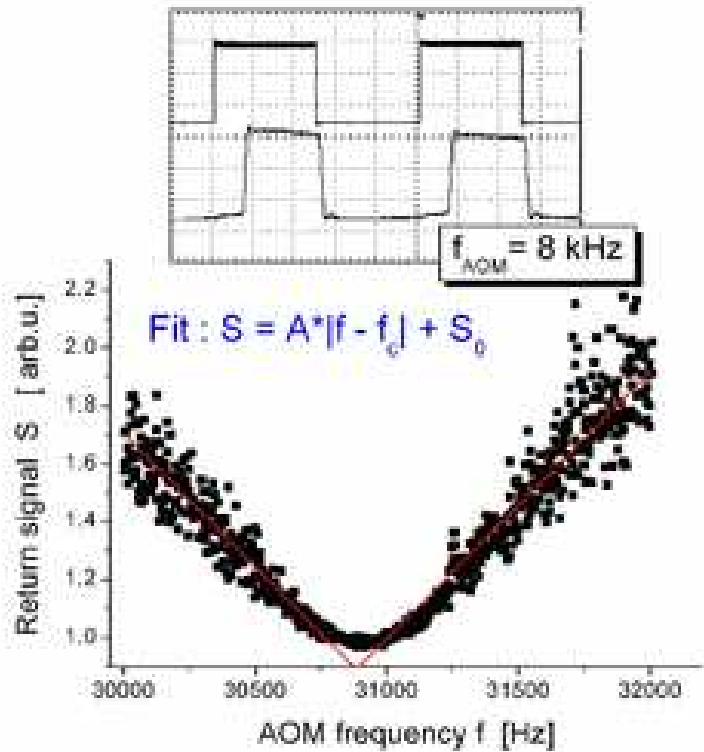}
\caption{The recorded signal of the photodiode. Upper part: AOM control signal and photodiode record for a fixed AOM frequency of 8~kHz; lower part: Collected light signal as a function of the AOM frequency.}
\label{fig.signal}\end{center}
\end{figure}

The results presented in this section have all been obtained by chopping the laser beam with an acousto-optic modulator.

The complete experimental set-up was automatically controlled by a LabView routine, allowing the variation of the AOM frequency in a well-defined frequency interval, and the recording and analysis of the corresponding photo-diode signal.

\subsection{Results}
A typical measurement result is shown in figure~\ref{fig.signal}.
The upper part of this figure shows the driving signal for the AOM (upper trace) as well as the recorded photodiode signal (lower trace). Note that the delay $\Delta t$ between these traces corresponds to the total traveling time of the light over distance $2d$, and that the driving and return signal simultaneously go to zero: when the AOM goes off, the return signal is blocked. Consequently, when the duration of the upper level $T_{on}=1/(2f)$ is exactly equal to $\Delta t$, the returning light is blocked exactly when it is arriving to the AOM, and there is no signal measured on the photodiode.

This is what one can observe on the lower part of figure~\ref{fig.signal}, where the photo-diode signal is plotted as a function of the AOM frequency.
The curve can be fitted by function $S(f) = A \times |f - f_c| + S_0$, with $S_0$ the signal off-set, $f_c$ the AOM's frequency corresponding to the minimum signal value of the curve, and $A$ the amplitude of the signal.

The most accurate value obtained for one night of experiment (typically 10 full frequency scans around $f_c$) led to a value of the speed of light $c_{Marseille} = 4d\times f_c =( 299,677 \pm 37 ) $~km/s, corresponding to a relative precision of $1.2\times10^{-4}$. For a refractive index of air evaluated as $n = 1.000296$, this value fits the expected speed of light with an accuracy of $9.3\times10^{-5}$.

\subsection{Systematic effects}

Various systematic effects have an influence on the deduced value of the speed of light. The incertitude of the distance measurement $d$ is the smallest contribution with a relative value of $1.35 \times 10^{-5}$ and can be neglected. An important perturbing factor is then certainly the fluctuation of the refractive index of air $n$. First, the difference of the measured value from the value of the speed of light in vacuum is clearly ruled by our estimation of the value of $n$. Second, and probably in an even more important manner, fluctuations of the refractive index in time and space influence the results~\cite{ams_n}. Excluding the set of measurements which have been made on one humid day, all series have been recorded with very similar meteorological conditions, which are a mean temperature of 299 K (at nightfall), in dry air at normal pressure (1015 hPa), which led to our evaluation of $n$.

Moreover, especially with the proximity to the Mediterranean sea, wind and dust are important perturbing factors as well. With changing wind conditions, the amplitude of the signal can vary up to a factor of ten due both to atmospheric perturbations and relative vibration of the measurement sites. Typically, one can see the signal blinking irregularly at frequencies of the order of a few hundred of Hz. Long integration times are then necessary to be able to achieve a good signal-to-noise ratio.

Finally, our precision is limited by the fitting of the data. As one can see on figure~\ref{fig.signal}, the part of the curve around the extinction is smooth and round instead of being extremely abrupt. Indeed, this is due to the bandwidth of the detection (1~GHz for the photodiode, but only 250~kHz for the acquisition card) that will induce a filtering of the very short transmitted impulsions and a distorsion of the signal. This artificial alteration of the shape of the curve makes it uneasy to fit with a basic calculation software.

Yet, given the experimental conditions, in an open space and in the presence of visitors, it is a fairly good result, even if it is still orders of magnitude less precise than the best values of $c$, obtained in the beginning of the 1970's~\cite{even72}.

\section{Outreach activities}
\label{sec.outreach}

Notre-Dame de la Garde lies in the heart of Marseille and its inhabitants. The laser beam which has been sent from the university site (close to the main train station) to a corner cube mounted on the church tower crosses the immediate city centre at a corner of the touristy harbor~\cite{autorisations}.

Three tents were installed during the experiment in very busy parts of town and almost under the laser beam. Explicative posters about the experiment, and also about the laser and its applications were accompanied by a small exhibition of some beautiful multi-color holograms, of several very visible and playful demonstration experiments, and of a reduced size model of the measurement set-up. The experiment was linked to the main tent via a webcam.

The experiment could also be followed on a blog~\cite{blog} relating the set-up and test details, as well as the day-to-day results along with many pictures and movies about the operation. On-site visits were made possible for small groups and individuals on registration, and attracted many visitors of all ages.

Due to Mie scattering the visibility and brilliance of the laser beam is high for a spectator on-axis of the beam, and the visibility decreases off-axis. Moreover, light scattered from different sources of street lighting may decrease the observed contrast along the beam strongly. The best view is therefore achieved directly from the experimental site, yet the beam was clearly visible to citizens all over town.

The described experiment has been fully accompanied by communication to the local media (TV and newspapers).  Our actions are continued throughout the year by "class ambassadors" visiting junior high schools and high schools, giving an introduction to the laser and its applications.

\section{Conclusion}

Our modern version of Fizeau's experiment is an extraordinary example of involving students and researchers of various levels and on divers topics, including lasers, optics, mechanics and electronics. This experiment can be realized with a low-power laser for the "measurement" part, involving all aspects of electronics, signal analysis and optical alignment and it can be set up in just a few weeks by a motivated master student. A spectacular visibility  can then be realized at night starting from 2~W of continuous power. Moreover, we noted the measurement of the speed of light is a very appealing notion to public, students and researchers at different levels. And that in any case, this bright green laser beam in the night sky certainly triggered the curiosity of all of those who had the chance to see it.

\section*{Acknowledgements}
This experiment would not  have been possible without the material
and technical input from different sources. SpectraPhysics has
provided us with the 10 W cw laser and Nicolas Treps has lent part of
the optical material which had been used in Paris in 2005. Our
activities have been financially supported by a large number of
local partners, and all the outreach activities have been made
possible by the help of almost fifty junior and senior physicists.

\section*{References}
\bibliographystyle{JphysB}


\begin{thebibliography}{}

\bibitem{amp2005} http://expositions.obspm.fr/lumiere2005/ \\ Jean-Louis Bobin, James Lequeux and Nicolas Treps, \textit{C'était "c" à Paris}, Bulletin de la SFP \textbf{ 152}, 32 (2005) 

\bibitem{fiz1849}
H Fizeau ; \textit{Sur une expérience relative à la vitesse de propagation de la lumière}, C.R. de l'Académie des Sciences, T.29 (1849) 90-92 and 132 (1849)

\bibitem{cornu}
His set-up and mesurement have though been improved by his scholar Alfred Cornu in 1872.\\ A. Cornu, \textit{Nouvelle determination de la vitesse de la lumiere},
[Recueil des travaux et discours d'A. Cornu parus de 1863 à 1904 dans des publications françaises et étrangères. Volume 2], Université Pierre et Marie Curie, Paris

\bibitem{roue}
http://lasersurmarseille.blogspot.com/search/label/Roue Fizeau

\bibitem{ams_n}
F. Fabry, C. Frush, I. Zawadzki, and A. Kilambi; \textit{On the Extraction of Near-Surface Index of Refraction Using Radar Phase Measurements from Ground Targets}, Journal of Atmospheric and Oceanic Technology, \textbf{14} 978-987 (1997)

\bibitem{even72}
K.M. Evenson, J.S. Wells, F.R. Petersen,B.L. Danielson, G.W. Day, R.L. Barger, R. L. and J.L. Hall, \textit{Speed of Light from Direct Frequency and Wavelength Measurements of the Methane-Stabilized Laser}, Phys. Rev. Lett. \textbf{29} 1346--1349 (1972)

\bibitem{autorisations}
In any case, permission has to be asked  for by the local authorities of civil aviation, even if the laser beam is sent almost horizontally.

\bibitem{blog}
http://www.lasersurmarseille.blogspot.com/


\end{thebibliography}

\end{document}